# Size-dependent intersubband optical properties of dome-shaped InAs/GaAs quantum dots with wetting layer

**Mohammad Sabaeian[*] and Ali Khaledi-Nasab**

*Physics Department, Faculty of Science, Shahid Chamran University, Ahvaz, Iran*

[*]*Corresponding author: sabaeian@scu.ac.ir*

In this work, the effect of size and wetting layer on subband electronic envelop functions, eigenenergies, linear and nonlinear absorption coefficients and refractive indices of a dome-shaped InAs/GaAs quantum dot were investigated. In our model, a dome of InAs quantum dot with its wetting layer embedded in a GaAs matrix was considered. A finite height barrier potential at the InAs/GaAs interface was assumed. To calculate envelop functions and eigenenergies, the effective one electronic band Hamiltonian and electron effective mass approximation were used. The linear and nonlinear optical properties were calculated by the density matrix formalism. © 2012 Optical Society of America

*OCIS codes:* 230.0230, 230.5590, 250.0250.

## 1. Introduction

Semiconductor quantum dot (QD) structures have attracted tremendous attention due to their unique physical properties and their potential applications in micro and optoelectronic devices [1-5] as well as life sciences and biotechnology [5-8]. Due to their relatively higher efficiency compared to bulk, they have also found applications as solar cell [5-12]. Performance improvements such as a threshold current independent of temperature, zero linewidth enhanced factor and extremely high differential gain have been achieved in QDs-based lasers [13]. They have also provided the possibility of generation of femtosecond pulses over a wide range of wavelengths [14].

In zero-dimension structures, the free carriers are confined to a small region by a so called confinement potential providing the quantization of electronic energy states based on the size of the dots. Atom-like discrete energy-levels are occurred when confining the carriers in a nano-region [15,16]. Photons with appropriate energy can cause the intersubband transitions involving large electric dipole moments [17,18]. The optical properties such as refractive index, absorption coefficient, and absorption cross section can be easily calculated once the linear and nonlinear susceptibilities of the QD are known. Large electric dipole matrix elements along with small energy differences between subbands, enhance the nonlinear contribution of dielectric constant, so one expects the light intensity plays a crucial role in the optical properties of the dots. In this regard the dot sizes can alter the values of electronic eigenenergies and their corresponding envelope functions.

Various shapes of QDs can be grown by Stranski-Krastanow (S-K) method [15,19-21]. The size and shape of such QDs depend on the growth conditions and the used techniques [22-24]. The optoelectronic properties of an ensemble of quantum dots are affected by the size distribution and the geometry of the dots [25-29].

The S-K method is essentially a self-organized hetero-epitaxial growth during molecular beam epitaxy (MBE) [10,15]. In this technique, after a number of lattice-mismatched atomic layers deposited on a barrier layer (substrate), accumulated strain energy forces transition from layer to island growth. This happens when a so called "wetting layer" reaches to a critical thickness of 3-4 nm [30-32]. The S-K approach however, leads to a broad size distribution with inherent [15].

The residual strain energy in dot structure can affect the energy eigenvalues, envelop functions and so the optical properties of QDs [22]. For example, Kim *et al.* observed a 90 nm blueshift in lowest-energy transition photoluminescence measurements for InGaS QDs grown on tensile-strained GaAsP compared with similar structures utilizing GaAs barriers [22]. Tansu *et al.* used GaAsP barriers surrounding the highly strained InGaAsN quantum well and realized highly performance lasers from 1170 nm up to 1400 nm wavelength regions [33]. We note InGaAsN material quantum wells have been realized to be high performance lasers by metalo-organic chemical vapor (MOCVD) and MBE methods [34]. However, for some material systems such as the well known InAs/GaAs system the strain doesn't play a dominate role [35]. Furthermore, the strain effects can be compensated using strain compensation layers as demonstrated by Nuntawog *et al.* [23,36]. The strain effect has also been compensated by Zhao *et al.* [37] and Park *et al.* [38] in InGaN/AlGaN and InGaN/InGaN quantum wells, respectively.

Recently, the MOCVD method has been used to grow nitride-based and arsenide-based QDs [12,39-42]. Kim *et al.* implemented this technique to growth InGaAs QDs on GaAsP matrix [22]. Ee *et al.* used MOCVD to grow InGaN quantum dots on GaN emitting at wavelength of 520nm [12]. Depending on In concentration, the band gap of InGaN QDs can cover the whole range of the visible spectrum [35].

Although self-assembled QDs are grown experimentally on wetting layers [43,44], most simulations excluded the wetting layer for simplicity in their calculations. In fact, the coupling between electronic states in the QD and wetting layer is essential to achieve better results [45-47]. Both localized QD and wetting-layer states must be considered as a coupled system.

Matthews *et al.* experimentally investigated the effect of wetting-layer on gain-current characteristics of InGaAs quantum dot laser [13]. They showed the population of wetting-layer states leads to a saturation of population inversion in QD states resulting gain saturation in QD laser. To confine the carriers in dot region, some authors used higher band-gap matrix embedding QD [48-50]. The result was a higher pick gain.

As we mentioned earlier, in the MBE and MOCVD methods the existence of wetting layer is essential to growth individual islands. In order to reduce the wetting layer effects and fully control the QDs formation, the selective area epitaxy has been employed [51-55]. In particular, utilizing the diblock copolymer lithography has gained highly uniform InGaN-based QDs with ultra-high density on nano-patterned GaN template [56].

Stier *et al.* purposed a model for the elastic, electronic, and linear optical properties of capped pyramid-shape InAs/GaAs DQs in the frame of eight-band **k.p** theory [25]. Li *et al.* computed the energy levels of an electron confined by disk-, ellipsoid- and conical-shaped InAs QDs embedded in GaAs [29]. Zhang *et al.* studied the wavelength shifts of CdSe QDs dispersed in a polymer host caused the quantum size effect and electro-optic Stark effect experimentally [57]. Seddik and Zorkani calculated the optical-absorption spectra of a hydrogenic donor impurity in a spherical CdSe QD with infinite potential confinement in the presence of magnetic field using variation and perturbation methods within the effective mass approximation [58]. Winkelnkemper *et al.* presented an eight-band $\mathbf{k.p}$ model to calculate the electronic properties of wurtzite $In_xGa_{1-x}N$/GaN QDs [35]. They considered the strain effects, piezoelectricity, pyroelectricity, spin-orbit and crystal-field splitting in their model without considering the wetting layer. Rostami *et al.* investigated the effect of centered defect on electrical and optical properties of spherical and cubic QDs, the former analytically and the latter numerically [59].

Liu and Xu obtained analytical expressions for envelop functions of an electron in a cylindrical QD [60]. They then calculated optical absorption coefficient and the variations in refractive index associated with intrasubband relaxation as a function of incident photon energy, dot size, and Al mole fraction $\beta$ in $Al_{\beta}Ga_{1-\beta}As$ material in the frame of density matrix formalism. Vahdani *et al.* investigated the effects of dot size and light intensity on optical absorption coefficient and refractive index changes of a parabolic quantum dot [61]. Xie studied the effect of laser on the hydrogenic impurity in linear and third order nonlinear absorption coefficient and refractive index change of a disc-like quantum dot [62]. Rezaei *et al.* first solved Schrodinger equation analytically to calculate envelope function and eigenenergies of a two-dimensional elliptic-shaped quantum dot [63]. They then used compact-density matrix formalism and iterative method to investigate the effect of size and optical light intensity on linear and nonlinear optical properties. Lu and Xie reported the impurity and exciton effects on linear and nonlinear optical properties of a disc-like quantum dot under a magnetic field [64]. He/she used one band effective mass theory to solve Schrodinger equation with appropriate Hamiltonian. Xie, then studied the nonlinear optical properties of a negative donor disk-like quantum dot with Gaussian confining potential [65]. Liang and Xie recently investigated the combined effects of hydrostatic pressure and temperature on the optical properties of a hydrogenic impurity in a disk-shaped QD in the presence of an external electric field [66].

In this paper, we report the dependency of the intersubband optical properties of dome-shaped InAs QDs embedded in GaAs matrix on dome radius and radiation intensity. A finite height potential barrier appropriate for a more realistic experimental situation is employed. For QDs with its wetting layer, the eigenenergies, envelop functions, electric dipole moment matrix components, relative linear and nonlinear refractive indices changes (RLRIC and RNRIC), and linear and nonlinear absorption coefficients (LAC and NAC) are all calculated versus the dome radius and light intensity. The wetting layer zone which is usually ignored in the previous articles [25,28,29,44-47,57-66] is fully adopted in this work and its effects on optoelectronic quantities are investigated in detail. Our model does not take the strain field effect into account. Although our approach is applied for III-V semiconductor optoelectronics devices, however it can be applied very well on other significant optoelectronic materials such as III-nitride semiconductors to generate short wavelengths. These materials that are usually used to generate 420-500 nm wavelengths, have been found applications for solid-state lighting and diode lasers, medicine, optical storage, terahertz photonics, power electronics, thermoelectricity and solar cells [67-73]. Also, our modeling can also be used for other systems using dilute nitride-based materials such as InGaAsN at 1.3-1.4 μm [33,34,74] and GaInNAsSb at 1.5μm [50,74-76] with high performance as diode lasers.

## 2. Envelop functions and eigenenergies of dome-shape QD

In the $\mathbf{k} \cdot \mathbf{r}$ approximation, the envelop function that modulates the periodic part of Bloch function at $\mathbf{k} = 0$, can describe the electrons wave function in QDs [77]. Consider a dome-shaped quantum dot with cylindrical symmetry whose Schrodinger wave equation in one band envelop function formalism is given by

$$-\frac{h^2}{8\pi^2 m^*}\nabla^2 \Psi(\vec{r}) + V(r)\Psi(\vec{r}) = E\Psi(\vec{r}) \qquad (1)$$

where $m^*$ is the electron effective mass having two different values in QD region and in surrounding medium, $\psi(\vec{r})$ is the electronic envelop function and E is the eigenenergy. Cylindrical symmetry allows us to use the separation of variables technique as:

$$\Psi(\vec{r}) = \chi(r,z)\Theta(\varphi) \qquad (2)$$

where $r$, $z$ and $\varphi$ are cylindrical coordinates. Substituting Eq.(2) in Eq.(1) and dividing both sides by $\chi(r,z)\Theta(\varphi)$ gives:

$$\frac{1}{\Theta}\frac{d\Theta}{d\varphi^2} = -l^2 \tag{3}$$

and

$$-\frac{m_e r^2 h}{8\pi^2}\frac{1}{\chi_l}\left[\frac{\partial}{\partial z}\left(\frac{1}{m_e}\frac{\partial \chi_l}{\partial z}\right) + \frac{1}{r}\frac{\partial}{\partial r}\left(\frac{r}{m_e}\frac{\partial \chi}{\partial r}\right)\right] + m_e r^2\ V - E\ = -\frac{h^2}{8\pi}l^2 \tag{4}$$

where $l$ is a separation constant. As the envelop function must be single-valued under $2\pi$ rotation, the $\varphi$-part can be expressed as $\Theta(\varphi) \propto \exp(il\varphi)$ leading to integer values of $l = 0, \pm 1, \pm 2,...$. Eq.(4) can be rearranged as

$$-\frac{h^2}{8\pi^2}\left[\frac{\partial}{\partial z}\left(\frac{1}{m_e}\frac{\partial \chi_l}{\partial z}\right) + \frac{1}{r}\frac{\partial}{\partial r}\left(\frac{r}{m_e}\frac{\partial \chi_l}{\partial r}\right)\right] + \left(\frac{h^2}{8\pi^2 m_e}\frac{1}{r^2} + V\right)\chi_l = E\chi_l \tag{5}$$

which has the general form of

$$\nabla.(-c\nabla\chi_l) + a\chi_l + \beta\nabla\chi_l = E_l\chi_l \tag{6}$$

with $\beta = \frac{-h^2}{8\pi^2 m_e}\frac{1}{r}$, $a = \frac{h^2}{\pi^2 m_e}\frac{l^2}{r^2} + V$ and $c = \frac{h^2}{8\pi^2 m_e}$.

## 2.1. Boundary conditions

To solve Eq. (6) and obtain the envelope functions and energy eigenvalues, certain boundary conditions are imposed. The simulation region and its boundaries are shown in Fig. (1). According to Fig.(1), for boundaries 1 (top) and 5 (bottom) the condition of $\chi_l = 0$ is considered, while for boundaries 2, 3, 4, 6, 7 and 8 the condition of $\mathbf{n}.(\nabla\chi_l) = 0$ is used in which $\hat{\mathbf{n}}$ is outward unit vector. For interface boundaries of InAs and GaAs, the condition of $\mathbf{n}.(\nabla \chi_l / m^*)_{\text{GaAs}} = \mathbf{n}.(\nabla \chi_l / m^*)_{\text{InAs}}$ is adopted because of the finiteness of the potential barrier.

## *2.2 Envelop functions and energy eigenvalues*

To solve the differential equation of (6) numerically, we adopted the finite element method (FEM). To carry out our simulations, we divided the whole simulation area into 7360 triangle elements. For one quarter of QD with its wetting layer 1040 elements and for upper region of QD and lower region of QD, 3504 and 2816 elements have been used, respectively. The effective electron mass was set to $0.023 m_e$ for InAs and $0.069 m_e$ for GaAs [45], where $m_e$ is the free electron mass. Also, the height of potential barrier was set as V=0.697 eV. The thickness of wetting layer corresponding to References [15,31] has been set as 3 nm in all our calculations. Figs. (2-a) to (2-f) show the ground and first excited state envelop functions for three values of dome radii of $r = 3\,nm$ (2-a and 2-b), $r = 7\,nm$ (2-c and 2-d) and $r = 15\,nm$ (2-e and 2-f). The results show localization of ground state envelope function in wetting layer for $r = 2\,nm$ to $r = 6\,nm$. With increasing the dome radius it then exits from the wetting layer and expands in dome region. Fig. (2-c) shows the ground state envelope function emerging from the wetting layer. For the excited state however this happens in two steps. At $r \approx 5\,nm$, one lobe exits, then at $r \approx 11\,nm$ the other lobe relaxes its overlap to wetting layer. This effect actually will impose its impact on electric dipole moments and related optical properties.

Fig. (3) shows the ground state (square) and first excited state (circle) energy eigenvalues versus the dome radius. For ground state, the eigenenergies are almost constant with dome radius increasing up to $r = 7\,nm$ where the corresponding envelope function starts emerging from the wetting layer. Eigenvalues curve then drops exponentially as dome radius increase. The similar exponentially decaying behavior is also seen for the first excited state eigenenergies showing two plateau segments which can be attributed to two steps emerging of envelope function from the wetting layer.

In Fig. (4), we plotted the eigenenergies difference, $\Delta E$, between the ground and the first excited states versus the dome radius; the quantity that is proportional to resonance frequency. The wetting layer role is obvious especially for radii less than $r \approx 11\,nm$.

## 3. Optical properties of dome-shape QD

### 3.1 Definitions

Let a linear x-polarized monochromatic electric field propagates along z-direction as:

$$\tilde{E}(z,t) = E_0 \hat{i}\, e^{i(kz-\omega t)} + C.C \tag{7}$$

where $k = n\omega/c$ is the complex propagation constant and $\omega$ is the angular frequency [16]. $N = n + in_i$ is the complex refractive index with its $n$ and $n_i$ as real and imaginary parts, respectively. The refractive index can be calculated through the effective susceptibility as

$$N = \sqrt{1 + \chi_{eff}(\omega)} \approx 1 + \frac{1}{2}\chi_{eff}(\omega) \tag{8}$$

where effective susceptibility is defined as $\chi_{eff}(\omega) = \chi^{(1)} + \chi^{(2)}(\omega)\tilde{E} + \chi^{(3)}(\omega)\tilde{E}^2$. $\chi^{(1)}$, $\chi^{(2)}$ and $\chi^{(3)}$ are the linear, second order and third order susceptibilities, respectively. The relative refractive index changes and absorption coefficient of the medium can be calculated as [61]:

$$\frac{\Delta n}{n} = \frac{n-1}{n} = \frac{1}{2}\mathrm{Re}\left(\frac{\chi_{eff}(\omega)}{n}\right) \tag{9}$$

and

$$\alpha = \frac{2n_i\omega}{c} = \omega\sqrt{\frac{\mu}{\epsilon_R}}\,\text{Im}\left[\epsilon_0 \chi_{\text{eff}}(\omega)\right] \tag{10}$$

respectively, where $\mu$ is the vacuum permeability and $\epsilon_R$ is the real part of permittivity. From Eqs. (9) and (10) one can conclude that the calculation of optical properties implies the calculation of linear and nonlinear susceptibilities.

The linear and nonlinear susceptibilities can be calculated through the quantum density matrix formalism [78]. According to this formalism, the linear susceptibility is calculated through the following formula:

$$\chi^{(1)} = \frac{\sigma}{\hbar\varepsilon_0} \frac{|M_{21}|^2}{\omega_{21} - \omega - i\gamma_{21}} \tag{11}$$

where $\sigma$ is the carrier density, $M_{21} = \langle \psi_2 | -ez | \psi_1 \rangle$ is the off-diagonal component of electric dipole moment matrix, $\omega_{21} = (E_2 - E_1)/\hbar$ is the transition angular frequency and $\gamma_{21}$ is the damping rate for off-diagonal elements of density matrix. Also, the second order susceptibility due to optical rectification and third order susceptibility are given respectively by [78]:

$$\chi_0^{(2)} = \frac{\sigma |M_{21}|^2}{\varepsilon_0 \hbar^2 \left[(\omega_{21} - \omega)^2 + \gamma_{12}^2\right]} \left\{ 2\gamma_{21}\left(\frac{M_{22}}{\gamma_{22}} - \frac{M_{11}}{\gamma_{11}}\right) \right. \\ \left. + \frac{2(M_{22} - M_{11})}{\omega_{21}^2 + \gamma_{21}^2}\left[\omega_{21}(\omega_{21} - \omega) - \gamma_{21}^2\right] \right\} \tag{12}$$

and

$$\chi^{(3)}(\omega) = -\frac{\sigma \, |M_{21}|^2}{\varepsilon_0 \hbar^3 \; \omega_{21} - \omega - i\gamma_{21}}$$
$$\times \left\{ \frac{4\,|M_{12}|^2}{\omega_{21} - \omega^2 + \gamma_{21}^2} - \frac{\left|M_{22} - M_{11}\right|^2}{\omega_{21} - i\gamma_{12} \; \omega_{21} - \omega - i\gamma_{12}} \right\}. \tag{13}$$

## 3.2 Results

In order to calculate the dipole moment matrices that are the major part of susceptibility calculations, we adopted the Simpson's double numerical integration method in a home-made code programmed in Maple. We set the carrier density and the damping rate to $\sigma = 3 \times 10^{22}\,\text{m}^{-3}$ and $\gamma_{21} = 5\,\text{ps}^{-1}$, respectively [16,61]. Also, the refractive index of GaAs was set to $n = 3.2$. According to Boyd notation [78], the relation between field intensity and electric field amplitude is expressed as $I = 2n\varepsilon_0 c |E|^2$.

Fig.(5) shows the absolute value of the diagonal elements of dipole moment matrix, i.e. $M_{11} = \left| \langle \psi_1 | -ez | \psi_1 \rangle \right|$ and $M_{22} = \left| \langle \psi_2 | -ez | \psi_2 \rangle \right|$, versus the dome radius. According to this figure, from $r = 2\,nm$ to $r = 6\,nm$, with increasing the dome radius, $M_{11}$ moment does not show noticeable change and remains approximately constant. This is due to totally localization of envelope function in wetting layer (see Fig. (2-a)). After $r = 6\,nm$, the ground state envelope function comes out from the wetting layer, so the dipole moment increases as the dome region expands. The increasing behavior of ground state dipole moment continues to bigger values of dome radius. However, for $M_{22}$ the increasing in its value takes place in two steps: from $r = 2\,nm$ to $r = 5\,nm$ where the excited state envelope function is totally localized in wetting

layer, $M_{22}$ is small. In the second step, according to Fig. (2-d), in the interval of $r = 6\,nm$ to $r = 10\,nm$, one lobe of two-parts excited state envelope function comes out from the wetting layer leading to a small rise in $M_{22}$. After $r = 10\,nm$, the excite state envelope function totally have come out from the wetting layer and so with increasing the dome radius and expanding the envelope function in dome region, $M_{22}$ increases. In the larger dome radii, $M_{22}$ is bigger than $M_{11}$.

Fig.(6) shows the off-diagonal element of dipole moment matrix, $M_{21}$, versus the dome radius revealing smaller values relative to $M_{11}$ and $M_{22}$ by one order of magnitude. It is clear that the behavior of $M_{12}$ is somehow more complicated than $M_{11}$ and $M_{12}$. This can be explained as follow: from $r = 2\,nm$ to approximately $r = 5\,nm$, the both ground and excited states lie in wetting layer. So, increasing the dome radius has not any noticeable effect on envelop functions. Then after $r = 5\,nm$ to $r = 8\,nm$, one part of excited state envelop function comes out from wetting layer whereas ground state does not. This decreases ground-first excited states overlapping and leads to a drop in $M_{12}$. From $r = 8\,nm$ to $r = 10\,nm$ where the ground state envelop function starts emerging from the wetting layer, $M_{12}$ increases. After $r = 10\,nm$ which both ground and first excited states exit completely from the wetting layer, the dipole moment grows with increasing the dome radius.

Fig. (7) shows the relative linear refractive index changes (RLRIC), $\Delta n^{(1)}/n$, as a function of photon energy for several dome radii. This quantity indeed does not depend on light intensity. From this figure we can see by increasing the dome radius, the RLRIC resonates at lower photon energies while its height is increasing. Fig. (8) shows the resonant peak heights of RLRIC versus the dome radius. This figure reveals an approximately constant value for RLRIC from $r = 2\,nm$ to $r = 10\,nm$, then a rapid change nearly $r = 11\,nm$ where both ground and first excited states exit from wetting layer. For bigger radii, the variations of resonance RLRIC versus the dome radius is approximately linear.

Fig. (9) shows the linear absorption coefficient (LAC), $\alpha^{(1)}$, as a function of photon energy for five dome radii bigger than $r = 10\,nm$. Similar to RLRIC case, the LAC does not depend on light intensity, too. By increasing the dome radius from the smaller values from $r = 10\,nm$, a dramatic change occurs for the resonant peak heights in the radial interval of $r = 10\,nm$ up to $r = 12\,nm$. This effect can be seen well from the Fig. (10) where we have plotted the height of LAC resonant peaks versus the dome radius. The variations of resonant LAC versus the radius, before $r = 10\,nm$ and after $r = 12\,nm$ are moderate.

Fig. (11) shows the variations of relative refractive index change due to second order optical rectification versus the photon energy. The solid curves denote $\Delta n^{(2)}/n$ for various dome radii of $r = 12nm$, $r = 14\,nm$, $r = 16\,nm$, $r = 18\,nm$ and $r = 20\,nm$. The light intensity is taken to be constant as $I = 0.1\,MW/cm^2$. The results corresponding to higher light intensities have been plotted by different style curves. The dashed, dash-dotted, and dotted curves stand for $I = 0.15\,MW/cm^2$, $I = 0.2\,MW/cm^2$, and $I = 0.3\,MW/cm^2$, respectively. From these curves it can be seen that the resonance frequency shifts toward the lower values when the dome radius increases. Moreover, the heights of resonance peaks are growing by light intensity increasing. Fig. (12) shows the height of the resonance peaks of $\Delta n^{(2)}/n$ versus the dome radius for several light intensities. According to these curves, for radius below $r = 10\,nm$, the peaks are approximately the same even for different light intensities, but above $10\,nm$ the peak heights variations are linear as a function of the dome radius. The curve slops become larger when the light intensity increases.

The optical rectification did not show any portion in absorption coefficient, due to purely reality of corresponding susceptibility, $\chi_0^{(2)}$.

Fig. (13) shows the variations of relative third order refractive index change (RTRIC) versus the photon energy. Similar to optical rectification case, to clarify the dome size and light intensity effects on RTRIC, different styled curves have been used. From the right to the left, solid curves show RTRIC for $r = 12\,nm$, $r = 14\,nm$, $r = 16\,nm$, $r = 18\,nm$, and $r = 20\,nm$ respectively, while the light intensity is taken to be constant as $I = 0.1\,MW/cm^2$. For other intensities, other styled curves have been used as follow: the dashed curves for $I = 0.15\,MW/cm^2$, the dash-dotted curves for $I = 0.2\,MW/cm^2$ and dotted curves for $I = 0.3\,MW/cm^2$. For each group, the resonance frequency shifts toward lower photon energies with increasing the dome radius. The height of resonant peaks grows by increasing the light intensity.

Fig. (14) shows the peaks heights of $\Delta n^{(3)}/n$ versus the dome radius for several light intensities. The curves behavior is similar to those of $\Delta n^{(2)}/n$ except the third order case is smaller by two orders of magnitude (see Fig. (12)).

Finally we investigated the dome size and light intensity effects on third order absorption coefficient (TAC). The results are presented in Figs. (15) and (16). In Fig. (15), we have plotted the TAC versus the photon energy for several intensities of $I = 0.1\,MW/cm^2$ (solid curves), $I = 0.15\,MW/cm^2$ (dashed curves), $I = 0.2\,MW/cm^2$ (dash-dotted curves) and $I = 0.3\,MW/cm^2$ (dotted curves). From the right to the left every group stands for radii of $r = 12\,nm$, $r = 14\,nm$, $r = 16\,nm$, $r = 18\,nm$ and $r = 20\,nm$, respectively. From these curves we see that the portion of third order in absorption coefficient is negative. The peaks heights versus the dome radius have been plotted in Fig. (16). The different styled curves stand for various intensities. Approximately above radius of $r = 12\,nm$, the variations of peaks height versus the dome radius have linear behavior.

## 4. Conclusion

In this work, firstly the effect of the size of a dome-shaped quantum dot on its energy eigenvalues, envelop functions and elements of dipole moment matrices were investigated. Secondly, the effects of the size and the light intensity on linear and nonlinear relative refractive indices and absorption coefficients were studied in detail. The effect of wetting layer which is usually ignored was pointed out for small radius domes. A jump in electronic and optical properties was seen once the dome radius becomes larger and the envelop function exits from the wetting layer. For the ground state envelop function beyond approximately $r \approx 5\,nm$ and for the first excited state one, beyond approximately $r \approx 10\,nm$ the corresponding eigenenergies show exponential decay with increasing dome radius. The elements of dipole moment matrices however show a linear behavior with dome radius.

For optical properties, the heights of the resonant peaks of the relative linear and nonlinear refractive indices changes, $\Delta n^{(1,2,3)}/n$, along with third order absorption coefficient (with its negative sign), show a linear variations versus the dome radius in all light intensities for radii above $r \approx 10\,nm$, whereas the linear absorption coefficient remains almost constant for all dome radii.

## Acknowledgments

The authors would like to thank Dr. I. Kazeminezhad (Shahid Chamran University) and Dr. M. R. K. Vahdai (Shiraz University) for their helpful discussions.

## Reference

1. H. Teleb, K. Abedi, and S. Golmohammadi, "Operation of quantum-dot semiconductor optical amplifiers under nonuniform current injection," Appl. Opt. **50**, 608-617 (2011).

2. A. Karimkhani and M. K. Moravvej-Farsh, "Temperature dependence of optical near field energy transfer rate between two quantum dots in nanophotonic devices," Appl. Opt. **49**, 1012-1019 (2010).

3. T. C. Newell, D. J. Bossert, A. Stintz, B. Fuchs, K. L. Malloy, and L. F. Lester, "Gain and Linewidth Enhancement Factor in InAs Quantum-Dot Laser Diodes," IEEE Quantum Electron. **11**, 1527-1529 (1999).

4. P. Bhattacharya, S. Ghosh, A. D. Stiff-Roberts, "Quantum dot optoelectronic devices," Annu. Rev. Mater. Res. **34**, 1-40 (2004).

5. K. Sun *et al.* "Applications of colloidal quantum dots," Microelectronics J. **40**, 644-649 (2009).

6. T. Jamieson *et al.* "Biological applications of quantum dots," Biomaterials **28**, 4717-4732 (2007).

7. B. L. Liang, Zh. M. Wang, Yu. I. Mazur and G. J. Salamo, "Photoluminescence of surface InAs quantum dot stacking on multilayer buried quantum dots" Appl. Phys. Lett. **89**, 243124 (2006).

8. P. A. S. Jorge, M. Mayeh, R. Benrashid, P. Caldas, J. L. Santos, and F. Farahi, "Applications of quantum dots in optical fiber luminescent oxygen sensors," Appl. Opt. **45**, 3760-3767 (2006).

9. Y. Zhou *et al.* "Efficiency enhancement for bulk-hetrojunction hybrid solar cells based on acid treated CdSe quantum dots and low bandgap polymer PCPDTBT," Solar Energy Materials & Solar Cells **95**, 1232-1237 (2011).

10. S. Suraprapapich, S. Thainoi, S. Kanjanachuchai, and S. Panyakeow, "Quantum dot integration in hetrostructure solar cell," Solar Energy & Solar Cells **90**, 2968-2974 (2006).

11. A. Luque, A. Marti, E. Antolin, and P. Garcia-Linares, "Intraband absorption for normal illumination in quantum dot intermediate band solar cells," Solar Energy Materials & Solar Cell **94**, 2032-2035 (2010).

12. Y.-Kh. Ee, H. Zhao, R. A. Arif, M. Jamil, and N. Tansu, "Self-assembled InGaN quantum dots on GaN emitting at 520nm grown by metalorganic vapor-phase epitaxy," J. Crystal Growth **310**, 2320-2325 (2008).

13. D. R. Matthews, H. D. Summers, P. M. Smowton, and M. Hopkinson, "Experimental investigation of the effect of wetting-layer states on the gain-current characteristics of quantum-dot lasers," Appl. Phys. Lett. **81**, 4904-4906 (2002).

14. E. U. Rafailov, P. Loza-Alvarez, W. Sibbett, G. S. Sokolovskii, D. A. Livshits, A. E. Zhukov, and V. M. Ustinov,"Amplification of femtosecond pulses over by 18 dB in a quantum-dot semiconductor optical amplifer," Photn. Tech. Lett. **15**, 1023-1025 (2003).

15. J. J. Coleman, J. D. Young and A. Garg, "Semiconductor quantum dot laser: A tutorial," J. Lightwave Tech. **29**, 499-510 (2011).

16. M. R. K. Vahdani and G. Rezaei, "Linear and nonlinear optical properties of a hydrogen donor in lens-shaped quantum dots," Phys. Lett. A, **373**, 3079-3084 (2009).

17. M. Barati, G. Rezaei and M. R. K. Vahdani, "Binding energy of a hydrogenic donor impurity in an ellipsoidal finite-potential quantum dot " Phys. Status Solidi B, **244**, 2605- (2007).

18. K. J. Kuhn, G. U. Lyengar, S. Yee, "Free carrier induced changes in the absorption and refractive index for intersubband optical transitions in $Al_xGa_{1-x}As/GaAs/Al_xGa_{1-x}As$ quantum wells," J. Appl. Phys., **70**, 5010 (1991).

19. V.-T. Rangel-Kuoppa, G. Chen, and W. Jantsch, "Electrical study of self-assembled Ge Quantum Dots in p-type Silicon. Temperature dependent Capacitance Voltage and DLTS study," Solid State Phenomena **178-179**, 67-71 (2011).

20. C. Lang, D. Nguen-Manh, and D. J. H. Cochayne, "Modelling Ge/Si quantum dot using finite element analysis and atomistic simulation," J. Phys. :Conference Series **29**, 141-144 (2006).

21. X. –F. Yang, K. Fu, W.-L. Xu, and Y. Fu, "Strain effect in determining the geometric shape of self-assembled quantum dot, "J. Phys. D: Appl. Phys. **42**, 125414 (2009).

22. N. H. Kim, P. Ramamurthy, L. J. Mawst, T. F. Kuech, P. Modak, T. J. Goodnough, D. V. Forbes, ans M. Kanshar, "Characteristics of InGaAs quantum dots grown on tensile-strained $GaAs_{1-x}P_x$ ," J. Appl. Phys. **97**, 093518 (2005).

23. N. Nuntawong, S. Birudavolu, C. P. Hains, H. Xu, and D. L. Huffaker, "Effect of strain compensation in staked 1.3μm InAs/GaS quantum dot active regions grown by metallographic chemical vapor deposition, "Appl. Phys. Lett. **85**, 3050-3052 (2004).

24. V. G. Dubrovskii *et al.* "Effect of growth kinetics on the structural and optical properties of quantum dot ensemble," J. Crystal Growth **276**, 47-59 (2004).

25. O. Stier, M. Grundmann and D. Bimberg, "Electronic and optical properties of strained QDs modeled by 8-band k.p theory," Phys. Rev. B, **59**, 5688-701 (1999).

26. D. Bimberg, M. Grundmann and N. N. Ledentsov, *Quantum Dot Hetrostructures*, (UK: Wiley).

27. I. Filikhin, E. Deyneka, G. Melikian, and B. Vlahovic, "Electron states of semiconductor quantum ring with geometry and size variations," Molecular Simulation, **31**, 779-785 (2005).

28. X. –F. Yang, X.-S. Chen, W. Lu and Y. Fu," Effects of shape and strain distribution of quantum dots on optical transmition in the quantum dot infrared photodetector," Nanoscale Res. Lett. **3**, 534-539 (2008).

29. Y. Li, Voskoboynikov, C. P. Lee, and S. M. Sze, "Computer simulation of electron energy level for different shape InAs/GaAs semiconductor quantum dots," Comput. Phys. Comm. **141**, 66-72 (2001).

30. D. Leonard, K. Pond, and P. M. Petroff, "Critical layer thickness for self-assembled InAs islands on GaAs," Phys, Rev. B **50**, 11687-11692 (1994).

31. T. Walther, A. G. Gullis, D. J. Norris, and M. Hopkinson, "Nature of the Stranski-Krastanow transition during epitaxy of InGaAs on GaAs," Phys. Rev. Lett. **86**, 2381-2384 (2001).

32. A. G. Gullis, D. J. Norris, T. Walther, M. A. Migliorato, and M. Hopkinson, "Stranski-Krastavow transition and epitaxial island growth," Phys. Rev. B **66**, 81305-81401 (2002).

33. N. Tansu, J.-Y. Yeh, and L. J. Mawst, "Physics and characteristics of high performance 1200 nm InGaAs and 1300-1400 nm InGaAsN quantum well lasers by metal-organic chemical vapor deposition," J. Phys. : Condens. Matter **16**, S3277-S3318 (2004).

34. N. Tansua and L. J. Mawst, "Current injection efficiency of InGaAsN quantum well lasers," J. Appl. Phys. **97**, 054502 (2005).

35. M. Winkelnkemper, A. Schliwa, and D. Bimberg, "Interrelation of structural and electronic properties in $In_xGa_{1-x}N$/GaN quantum dots using an eight-band k.p model," Phys. Rev. Lett. **74**, 1553222 (2006).

36. N. Nuntawong, J. Tatebayashi, P. S. Wong, and D. L. Huffaker, "Localized strain reduction in strain-compensated InAs/GaAs stacked quantum dot structure," Appl. Phys. Lett. **90**, 163121 (2007).

37. H. Zhao, R. A. Arif, Y. K. Ee, N. Tansu, "Self-consistent analysis of strain-compensated InGaN-AlGaN quantum wells for laser and light emitting diodes, " IEEE J. Quantum Electron. **45**, 66 (2009).

38. S. H. Park, Y. T. Moon, J. S. Lee, H. K. Kwon, J. S. Park, and D. Ahn, "Spontaneous emission rate of green strain-compensated InGaN/InGaN LEDs using InGaN substrate, Phys. Status Solidi A **208**, 195 (2011).

39. D. Simeonov, E. Feltin, J. F. Carlin, R. Butte, M. Ilegems, N. Grandjean, " Stranski-Kranstanov GaN/AlN quantum dots grown by metal organic vapor phase epitaxy," J. Appl. Phys. **99**, 083509 (2006).

40. S. Ruffenach, B. Maleyre, O. Briot, B. Gil, " Grown of InN quantum dots by MOVPE, " Phys. Status Solidi C **2**, 826 (2005).

41. H. Y. Liu, S. L. Liew, T. Badcock, D. J. Mowbray, M. S. Skolnick, S.K. Ray, T. L. Choi, K. M. Groom, B. Stevens, F. Hasbullah, C. Y. Jim, M. Hopkinson, R.A.Hogg, "P-doped 1.3 µm InAs/GaAs quantum-dot lasers with a low threshold current density and high differential efficiency, " Appl. Phys. Lett. **89**, 073113 (2006).

42. R. L. Sellin, C. Ribbat, M. Grundmann, N. N. Ledentsov, D. Bimberg, "close-to-ideal device characteristics of high-power InGaAs/GaAs quantum dot laser," Appl. Phys. Lett. **78**, 1207 (2001).

43. M. S. Skolnick *et al.*, "Electronic structure of InAs/GaAs self-assembled quantum dots studied by perturbation spectroscopy," Physica E **6**, 348-357 (2000).

44. R. Oshima, N. Kurihara, H. Shigekawa and Y. Okada, "Electronic states of self-organized InGaAs quantum dots on GaAs (3 1 1) B studied by conductive scanning probe microscopy, " Physica E **21**, 414-418 (2004).

45. R. V. N. Melnik and K. N. Zotsenko, "Finite element analysis of coupled electronic states in quantum dot nanostructures," Modelling Simul. Mater. Sci. Eng. **12**, 465-477 (2004).

46. F. Adeler *et al.*, "Optical transition and carrier relaxation in self-assembled InAs/GaAs quantum dots," J. Appl. Phys. **80**, 4019-26 (1996).

47. R. V. N. Melnik and M. Willatzen, "Modelling coupled motion of electrons in quantum dots with wetting layers," Proc. Modelling and Simulation of Micrisystems (MSM) Conf. (USA, 21-25 April 2002), 506-509 (2002).

48. D. Colombo, *et al.* "Efficient room temperature carrier trapping in quantum dots tailoring the wetting layer,"J. Appl. Phys. **94**, 6513 (2003).

49. J. Jiang *et al.*, "High detectivity InGaAs/InGaP quantum-dot infrared photodetectors grown by low pressure metalorganic chemical vapor deposition," Appl. Phys. Lett. **84**, 2166 (2004).

50. J. S. Kim *et al.* , "Effects of high potential barrier on InAs quantum dots and wetting layer," J. Appl. Phys. **91**, 5055 (2002).

51. M. Helfrich, R. Groger, A. Forste, D. Litvinov, D. Gerthsen, T. Schimmel, D. M. Schaadt, "Investigation of Pre-structured GaAs Surfaces for Subsequent Site-selective InAs Quantum Dot Growth," Nanoscale Res. Lett. **6**, 211 (2011).

52. R. R. Li, P. D. Dapkus, M. E. Thompson, W. G. Jeong, C. Harrison, P. M. Chaikin, R. A. Register, D. H. Adamson, "Dense Arrays of Ordered GaAs Nanostructures by Selective Area Growth on Substrates Patterned by Block Copolymer Lithography," Appl. Phys. Lett. **76**, 1689 (2000).

53. K. Tachibana, T. Someya, S. Ishida, Y. Arakawa, "Selective Growth of InGaN Quantum Dot Structures and Their Microphotoluminescence at Room Temperature," Appl. Phys. Lett. **76**, 3213 (2000).

54. T. F. Kuech, L. J. Mawst, "Nanofabrication of III-V Semiconductors Employing Diblock Copolymer Lithography," J. Phys. D Appl. Phys. **43**, 183001 (2010).

55. J. H. Park, J. Kirch, L. J. Mawst, C.-C. Liu, P. F. Nealley, and T. F. Kuech, "Controlled growth of InGaAs/InGaAsP quantum dots on InP substrates employing diblock copolymer lithography, " Appl. Phys. Lett. **95**, 113111 (2009).

56. G. Liu, H. Zhao, J. H. Park, L. J. Mawst, and N. Tansu, "Selective area epitaxy of ultra-high density InGaN quantum dots by diblock copolymer lithography," Nanoscale Res. Lett. **6**, 342 (2011).

57. F. Zhang, L. Zhang, Y-X. Wang and R. Claus, "Enhanced absorption and electro-optic Pockels effect of electrostatically self-assembled CdSe quantum dots," Appl. Opt. **44**, 3969-3976 (2005).

58. A. D. Seddik and I. Zorkani, "Optical properties of a magneto-donor in a quantum dot," Physica E **28**, 339-346 (2005).

59. A. Rostami, H. Rasooli Saghai, N. Sadoogi, and H. Baghban Asghari Nejad, "Proposal for ultra-high performance infrared Quantum Dot," Opt. Express **16**, 2752-2763 (2008).

60. C. H. Liu and B-R. Xu, "Theoretical study of the optical absorption and refraction index change in a cylindrical quantum dot," Phys. Lett. A **372**, 888-892 (2008).

61. M. R. K. Vahdani and G. Rezaei, "Intersubband optical properties absorption coefficients and refractive index changes in a parabolic cylinder quantum dot," Phys. Lett. A, **374**, 637-643 (2010).
23

62. W. Xie, "Laser radiation effects on optical absorptions and refractive index in a quantum dot," Opt. Commun. **283**, 3703-3706 (2010).

63. G. Rezaei, Z. Mousazadeh, and B. Veseghi, "Nonlinear optical properties of a two dimensional elliptic quantum dot," Physica E **42**, 1477-1481 (2010).

64. L. Lu and W. Xie, "Impurity and exciton effects on the nonlinear optical properties of a disc-like quantum dot under a magnetic field," Superlattices and Microstructures **50**, 40-49 (2011).

65. W. Xie, "A study of nonlinear optical properties of a negative donor quantum dot," Opt. Commun. **284**, 4756-4760 (2011).

66. Sh. Liang and W. Xie, "Effects of the hydrogenic pressure and temperature on optical properties of a hydrogenic impurity in the disc-like quantum dot," Physica B **406**, 2224-2230 (2011).

67. H. Zhao, J. Zhang, G. Liu, and N. Tansu, "Surface Plasmon dispersion engineering via double-metallic Au/Ag layers for III-nitride based light-emitting diodes," Appl. Phys. Lett. **98**, 151115 (2011).

68. S.-H. Park, D. Ahn, J. Park, and Y.-T. Lee, "Optical properties of staggered InGaN/InGaN/GaN quantum well structures with Ga-and N-faces," Jap. J. Appl. Phys. **50**, 072101 (2011).

69. H. Zhao, G. Liu, J. Zhang, J. D. Poplawsky, V. Dierolf, and N. Tansu, "Approaches for high internal quantum efficiency green InGaN light-emitting diodes with large overlap quantum wells," Opt. Express **19**, A991-A1007 (2011).

70. R. M. Farrell, D. A. Haeger, P. S. Hsu, M. C. Schmidt, K. Fugito *et al.*, "High-power blue-violet AlGaN-cladding-free m-plane InGaN/GaN laser diodes, "Appl. Phys. Lett. **99**, 171113 (2011).

71. J. Zhang and N. Tansu, "Improvement in spontaneous emission rates for InGaN quantum wells on ternary InGaN substrate for light-emitting diodes," J. Appl. Phys. **110**, 113110 (2011).

72. Y. Li, B. Liu, R. Zhang, Z. Xie, and Y. Zheng, "Investigation of optical properties of InGaN-InN-InGaN/GaN quantum-well in the green spectral regime," Physica E **44**, 821-825 (2012).

73. J. Zhang, H. Zhao, and N. Tansu, "Large optical gain AlGaN-GaN quantum wells laser active region in mid-and deep-ultraviolet spectral regimes," Appl. Phys. Lett. **98**, 171111 (2011).

74. J. W. Ferguson, P. Blood, P. M. Smowton, H. Bae, T. Sarmiento, J. S. Harris, N. Tansu, and L. J. Mawst, "Optical gain in GaInNAs and GaInNAsSb quantum well," IEEE J. Quantum Electron. **47**, 870-877 (2011).

75. S. R. Bank, M. A. Wistey, H. B. Yuen, L. L. Goddard, W. Ha and J. S. Haris, "Low-threshold CW GaInNAsSb/GaAs laser at 1.49 μm," Electronics Lett. **39**, 1455 (2003).

76. S. R. Bank, L. L. Goddard, M. A. Wistey, H. B. Yuen, and J. S. Harris, "On the temperature sensitivity of 1.5 μm GaInNAsSb lasers," IEEE J. Sel. Top Quantum Electron. **11**, 1089-1098 (2005).

77. S. Datta, Quantum Phenomena, Addison-Wesley, Modular Series of Solid-State devices, (1989).

78. R. Boyd, *Nonlinear Optics*, 3rd Edition.

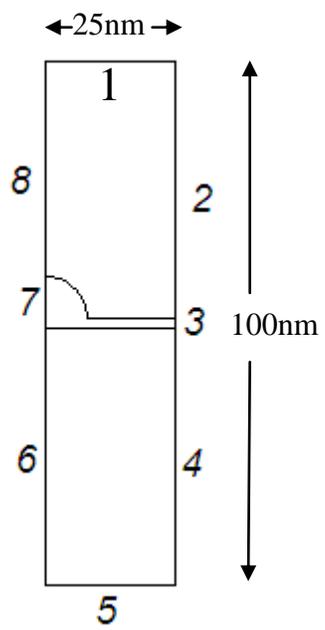

Fig. 1: The simulation area with numbered boundaries. The thickness of wetting layer has been set as 3 nm.

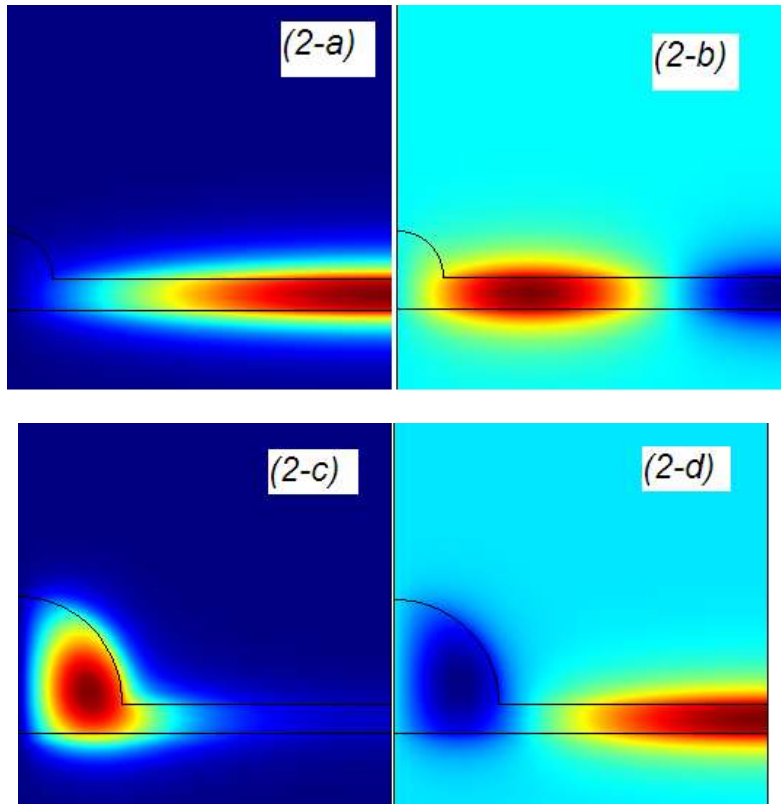

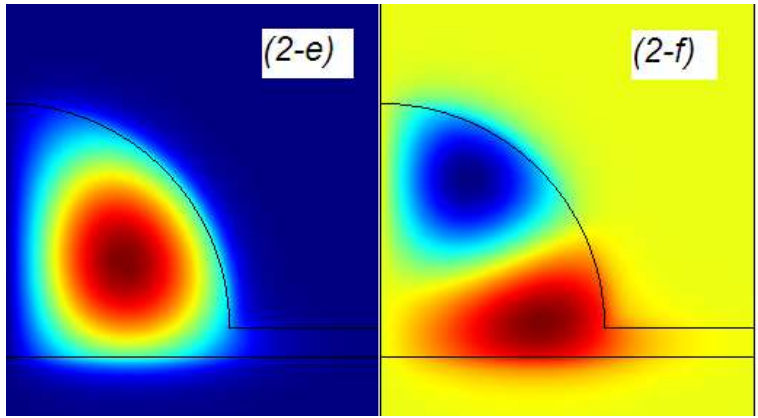

Fig. 2. The normalized ground state (left) and first excited state (right) envelop functions for r=3nm (2-a and 2-b), r=7nm (2-c and 2-d) and r=15nm (2-e and 2-f).

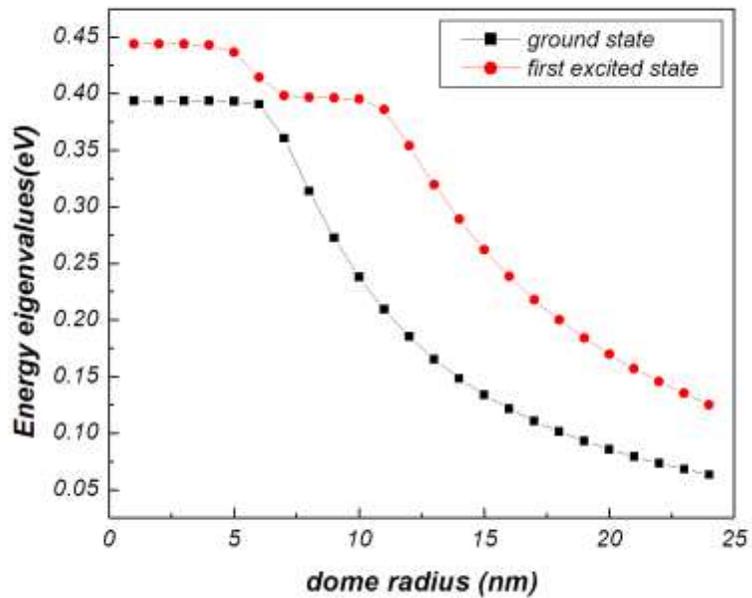

Fig. 3. The ground state (cubic) and the first excited state (circle) energy eigenvalues against the dome radius.

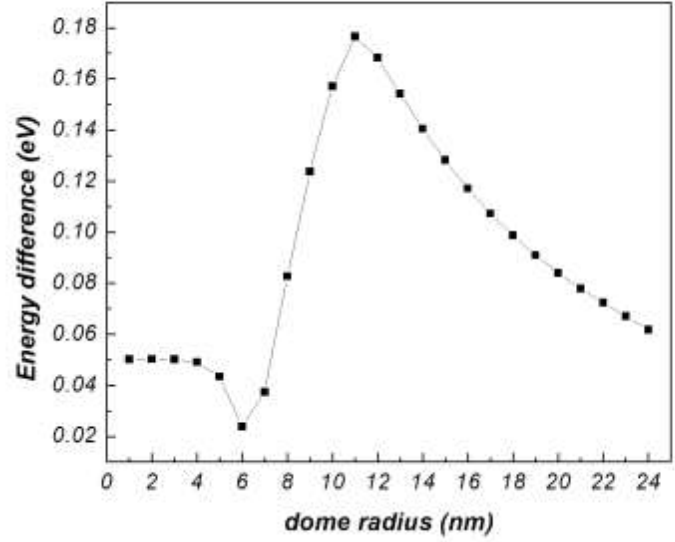

Fig. 4. The difference between ground and first excited state energies against the dome radius.

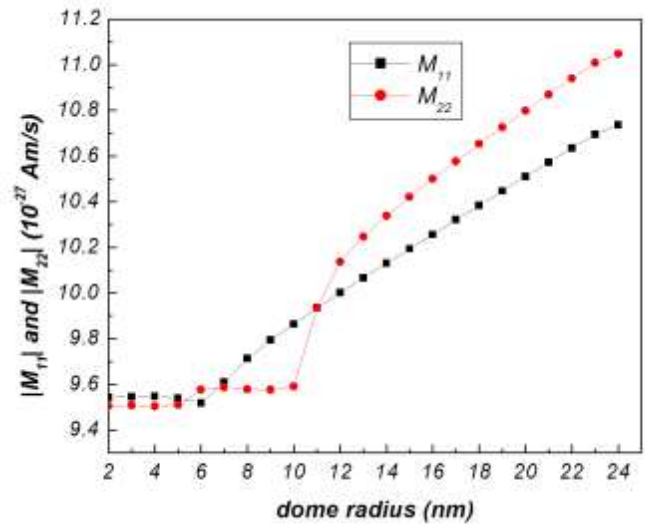

Fig. 5. The absolute value of diagonal elements of dipole moment matrix, $M_{11}$ (cube) and $M_{22}$ (circle), against the dome radius.

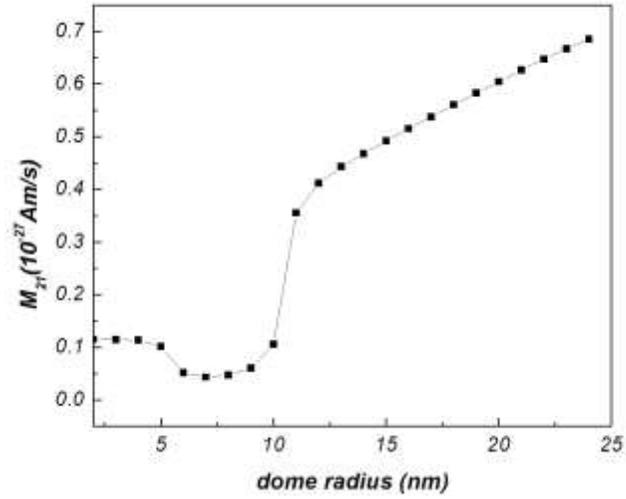

Fig. 6. The off-diagonal dipole moment element, $M_{21}$, against the dome radius.

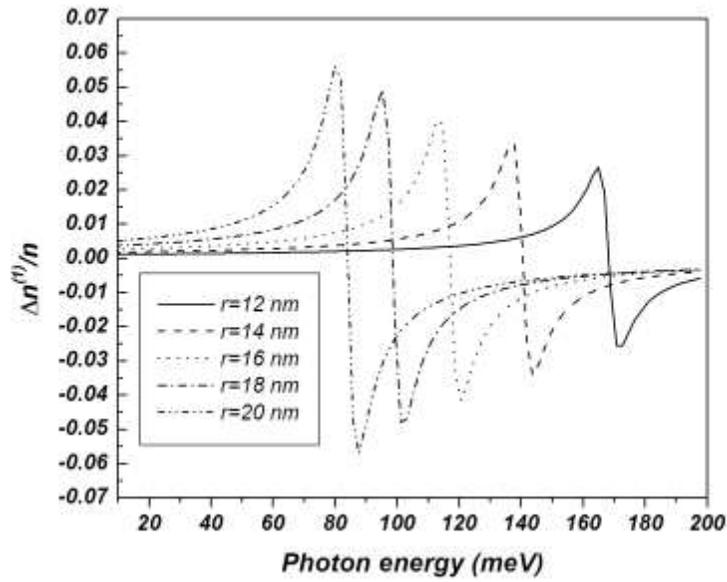

Fig. 7. The RLRIC versus the photon energy for various dome radii of $r = 12\,nm$, $r = 14\,nm$, $r = 16\,nm$, $r = 18\,nm$ and $r = 20\,nm$.

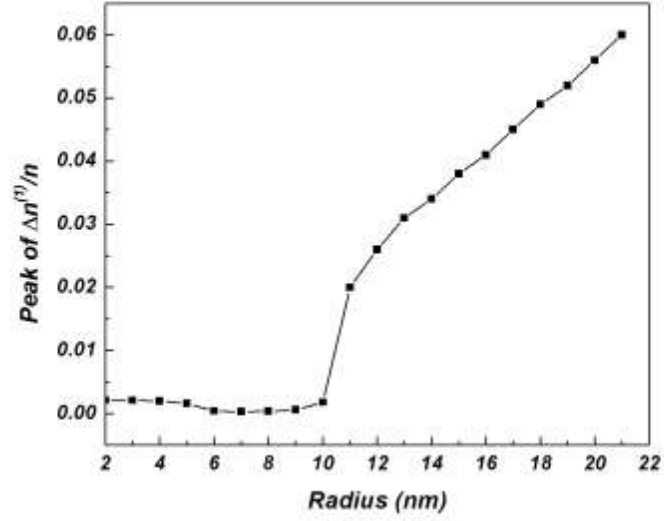

Fig. 8. The heights of resonate peaks of RLRIC, $\Delta n^{(1)}(\omega = \omega_{21})/n$, versus the dome radius.

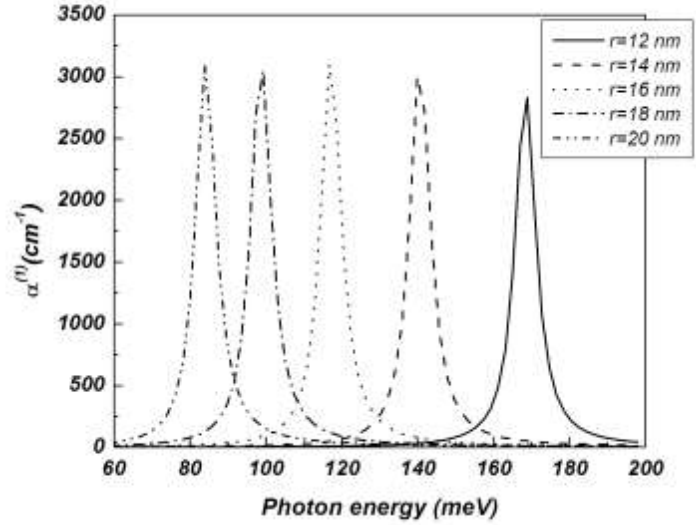

Fig. 9. The LAC versus the photon energy for various dome radii of $r = 12\,nm$ (solid), $r = 14\,nm$ (dashed), $r = 16\,nm$ (dotted), $r = 18\,nm$ (dashed-dotted) and $r = 20\,nm$ (dashed-dotted-dotted).

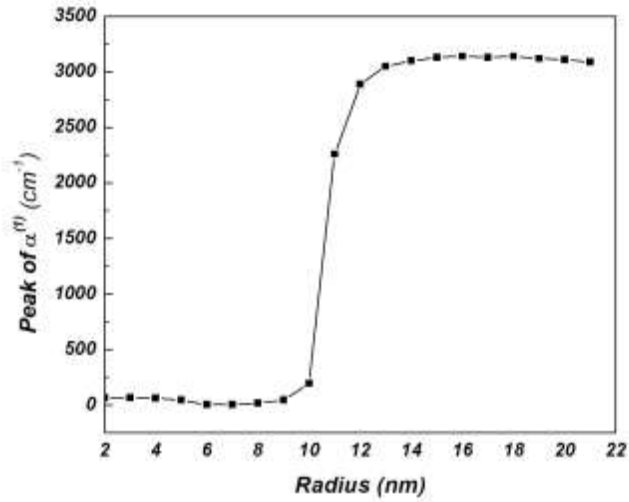

Fig. 10. The heights of resonant peaks of LAC versus the dome radius.

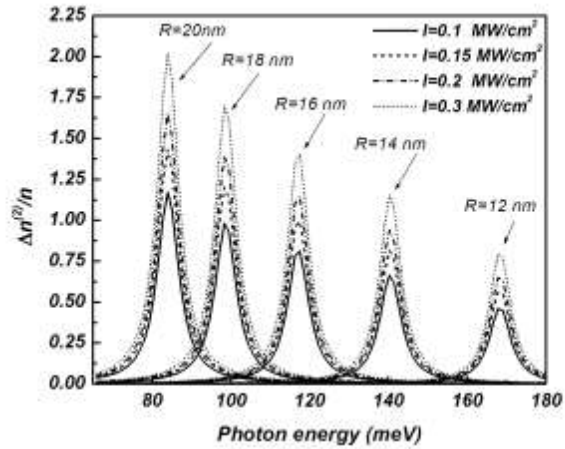

Fig. 11. The relative refractive index change due to optical rectification versus photon energy for several light intensities of $I = 0.1\,MW/cm^2$ (solid curves), $I = 0.15\,MW/cm^2$ (dashed curves), $I = 0.2\,MW/cm^2$ (dash-dotted curves) and $I = 0.3\,MW/cm^2$ (dotted curves).

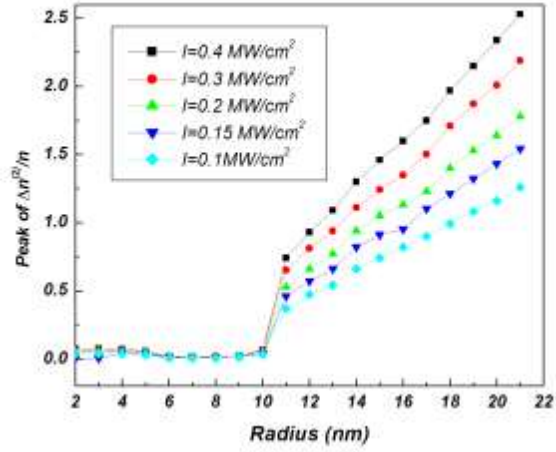

Fig. 12. The relative refractive index change due to optical rectification at resonate frequency, $\Delta n^{(2)}(\omega = \omega_{21})/n$, versus dome radius for various intensities.

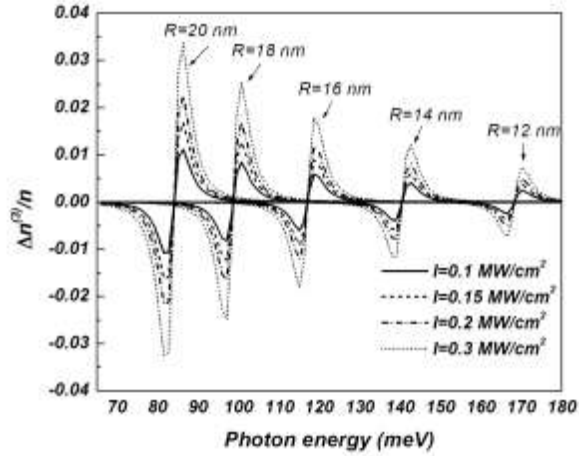

Fig. 13. RTRIC against the versus photon energy for several light intensities of $I = 0.1\,MW/cm^2$ (solid curves), $I = 0.15\,MW/cm^2$ (dashed curves), $I = 0.2\,MW/cm^2$ (dash-dotted curves) and $I = 0.3\,MW/cm^2$ (dashed curves). From the right to the left, every group of curves stands for dome radius of $r = 12\,nm$, $r = 14\,nm$, $r = 16\,nm$, $r = 18\,nm$ and $r = 20\,nm$.

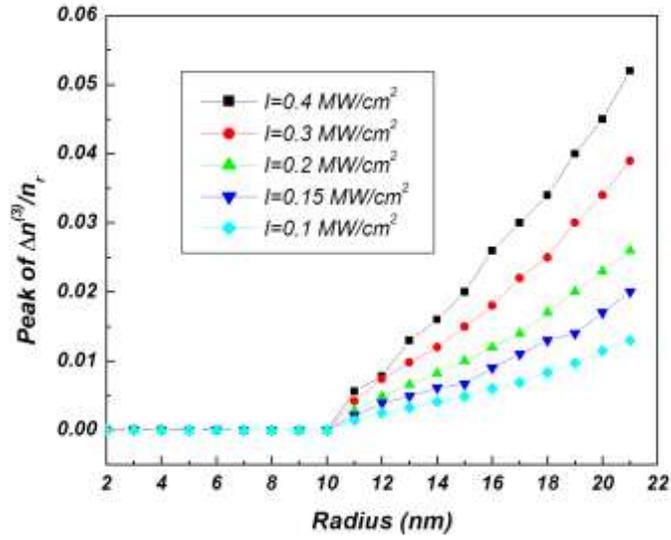

Fig. 14. The RTRIC at resonance frequency, $\Delta n^{(3)}(\omega = \omega_{21})/n$, versus the dome radius for various intensities.

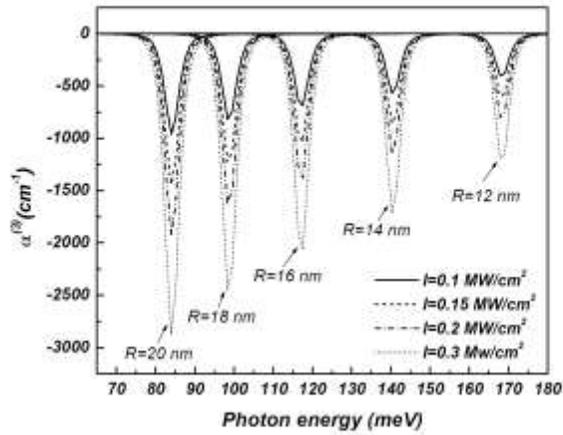

Fig.15. TAC against the photon energy for several light intensities of $I = 0.1\,MW/cm^2$ (solid curves), $I = 0.15\,MW/cm^2$ (dashed curves), $I = 0.2\,MW/cm^2$ (dash-dotted curves) and $I = 0.3\,MW/cm^2$ (dotted curves) and several dome radii. From right to the left: $r = 12\,nm$, $r = 14\,nm$, $r = 16\,nm$, $r = 18\,nm$ and $r = 20\,nm$.

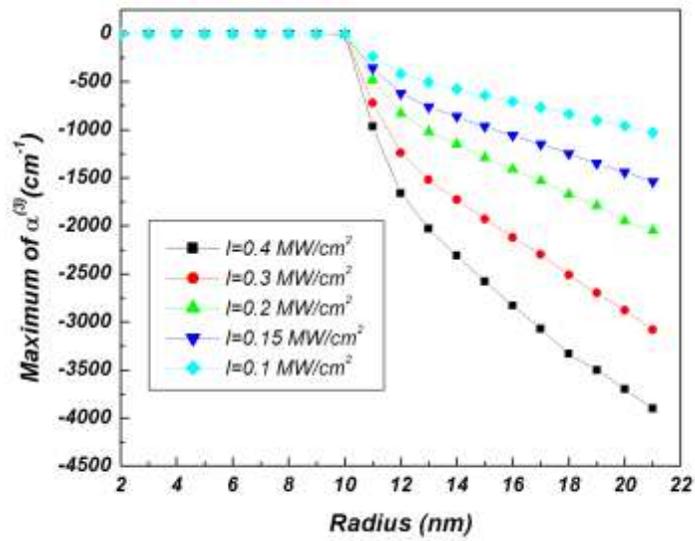

Fig.16. The TAC at resonante frequency $\alpha^{(3)}(\omega = \omega_{21})$ versus the dome radius for various intensities.